\documentclass[useAMS,usenatbib,usegraphicx]{mn2e}

\title[The origin of metals in DZ stars]{Rocky planetesimals as the origin of metals in DZ stars}

\author[J. Farihi et al.]{J. Farihi$^1$, M. A. Barstow$^1$, S. Redfield$^2$, P. Dufour$^3$, 
	N. C. Hambly$^4$\\
$^1$Department of Physics \& Astronomy, University of Leicester, Leicester LE1 7RH, UK; 
	jf123@star.le.ac.uk\\
$^2$Astronomy Department, Van Vleck Observatory, Wesleyan University, Middletown, CT 
	06459, USA\\
$^3$D\'epartement de Physique, Universit\'e de Montr\'eal, Montr\'eal, QC H3C 3J7, Canada\\
$^4$Institute for Astronomy, University of Edinburgh, Royal Observatory, Edinburgh EH9 3HJ, 
	UK}

\begin{document}

\date{}

\maketitle

\label{firstpage}

\begin{abstract}
The calcium and hydrogen abundances, Galactic positions and kinematics of 146 DZ 
white dwarfs from the Sloan Digital Sky Survey are analyzed to constrain the possible 
origin of their externally polluted atmospheres.  There are no correlations found between 
their accreted calcium abundances and spatial - kinematical distributions relative to 
interstellar material.  Furthermore, two thirds of the stars are currently located above 
the Galactic gas and dust layer, and their kinematics indicate multi-Myr residences in 
this region where interstellar material is virtually absent. 

Where detected, the hydrogen abundances for 37 DZA stars show little or no correlation 
with accreted calcium or spatial - kinematical distributions, though there is a general trend
with cooling age.  It is found that Eddington type accretion of interstellar hydrogen can 
reproduce the observed hydrogen abundances, yet simultaneously fails to account for 
calcium.  The calcium-to-hydrogen ratios for the DZA stars are dominated by super-solar 
values, as are the lower limits for the remaining 109 DZ stars.  All together, these polluted 
white dwarfs currently contain 10$^{20\pm2}$\,g of calcium in their convective envelopes, 
commensurate with the masses of calcium inferred for large asteroids.

A census of current $T_{\rm eff}\la12\,000$\,K, helium-rich stars from the Sloan Digital 
Sky Survey suggests the DZ and DC white dwarfs belong to the same stellar population, 
with similar basic atmospheric compositions, effective temperatures, spatial distributions, 
and Galactic space velocities.  Based on this result, pollution by the interstellar medium 
cannot simultaneously account for both the polluted and non-polluted sub-populations.  
Rather, it is probable that these white dwarfs are contaminated by circumstellar matter; 
the rocky remains of terrestrial planetary systems.  

In this picture, two predictions emerge.  First, at least 3.5\% of all white dwarfs harbor the 
remnants of terrestrial planetary systems; this is a concrete lower limit and the true fraction 
is almost certainly, and perhaps significantly, higher.  Therefore, one can infer that at least
3.5\% of main-sequence A- and F-type stars build terrestrial planets.  Second, the DZA 
stars are externally polluted by both metals and hydrogen, and hence constrain the 
frequency and mass of water-rich, extrasolar planetesimals.
\end{abstract}


\section{INTRODUCTION}

The curious presence of heavy elements in the atmospheres of cool white dwarfs has been 
known since the discovery of the first degenerate stars; van Maanen 2 is the prototype DZ white 
dwarf \citep{van17}.  For more than half a century it has been understood that, in the absence 
of radiative forces, elements heavier than hydrogen and helium should rapidly sink below the 
photosphere in the high gravity environment of white dwarfs \citep{sch48}; hence metal-lined 
white dwarfs like vMa 2 must be externally polluted.

Owing to a relatively low atmospheric opacity, the detection of calcium in the optical spectra 
of more than one dozen cool, single, helium-rich (DZ) white dwarfs preceded the availability 
of 8\,m class telescopes and high resolution CCD spectrographs \citep{sio90b}.  Although the 
sinking timescales are always orders of magnitude shorter than the white dwarf cooling age, 
heavy elements can persist for up to $10^6$\,yr in the sizable convection zones of helium-rich 
degenerates \citep{paq86}.  This fact led somewhat naturally to the hypothesis that the source 
of the accreted heavy elements was the interstellar medium (ISM), encountered in sufficient 
density a few to several times per Galactic orbit (on average) for the Solar neighborhood 
\citep{dup93a,dup93b,dup92}.

Several factors have continually challenged the ISM pollution scenario.  First, the typical 
dearth of hydrogen relative to calcium in DZ stars argues for the accretion of volatile depleted 
material \citep{duf07,wol02,sio90a}.  Second, the firm establishment of the hydrogen-rich DAZ 
spectral class, with heavy element diffusion timescales as short as a few days, and Galactic 
positions far from known interstellar clouds \citep{koe05a,zuc03}.  Third, the increasing number 
of metal-contaminated white dwarfs with infrared excesses and closely orbiting rings of dusty 
and gaseous debris \citep{far09a,jur09a,gan08}.  Thus, while there is firm observational 
evidence of pollution via circumstellar material, as yet there is none favoring the ISM.

\citet{aan93} provided an extensive spatial, kinematical, and calcium abundance analysis 
against which they tested likely scenarios of interstellar accretion based on the best available
data at that time, concluding interstellar accretion could not explain the abundances in {\em 
any} of their 15 DZ white dwarfs.  However, their landmark study was limited to targets within 
50\,pc for the most part, and to stars that were proper-motion selected, and hence biased 
toward higher space velocities \citep{sio88}.  \citet{kil07} examined possible correlations
between the current positions of 35 DAZ white dwarfs and models of the surrounding ISM
as inferred from nearby column density measurements, finding a lack of sufficient material
to account for most polluted stars.

This paper returns to the question of the origin of the DZ spectral class.  Historically, the
DZ stars outnumbered their hydrogen-rich counterparts stars by 25:1 \citep{dup93b} until 
high resolution, spectroscopic searches with large telescopes uncovered a commensurate 
number of DAZ stars \citep{koe05a,zuc03}.  Today, the ratio of known DZ to DAZ stars is 
greater than 4:1, due almost exclusively to the Sloan Digital Sky Survey (SDSS; \citealt{eis06}).  
The ability to detect mild levels of heavy element pollution in these stars, combined with the 
relative lack of circumstellar material at both the DZ and cool DAZ white dwarfs \citep{far09a}, 
makes the origin of the DZ stars challenging to constrain.

It is argued here that the hundreds of newly identified, cool, helium-rich white dwarfs from 
the SDSS provide an excellent, spatially and kinematically unbiased sample upon which to 
test the ISM accretion hypothesis.  These stars form a large statistical sample that favors an 
alternative to the accretion of Galactic gas and dust; as a whole the SDSS DZ stars do not 
show the spatial, kinematical, or elemental abundance correlations expected from accretion 
episodes within the plane of the Galaxy.  Furthermore, the SDSS DZ and DC (helium-rich, 
featureless) stars appear to be similar populations of polluted and non-polluted stars broadly 
sharing all relevant characteristics save photospheric calcium, and perhaps hydrogen.

\section{COOL HELIUM-RICH WHITE DWARFS FROM THE SDSS DR4}

The SDSS fourth data release (DR4; \citealt{ade06}) contains a prodigious number of new
white dwarfs, in the neighborhood of 6000 \citep{eis06}.  Excluding multiple entries for the 
same source, binaries or binary candidates, and uncertain classifications, there are 95 DZ 
and 142 DC stars within the DR4 white dwarf catalog.  Under further scrutiny, \citet{duf07} has 
expanded the number of confirmed, unique DZ stars to 146, providing effective temperatures, 
calcium abundances, and photometric distances under the assumption of $\log\,[g\,({\rm 
cm\,s}^{-2})]=8.0$.  Of these 288 cool, helium-rich white dwarfs in DR4, only one was known 
previously; G111-54 or SDSS J080537.64$+$383212.4 \citep{duf07,gre75}.

Perhaps surprisingly, the current number of DC and DZ stars in the SDSS are nearly identical.
The DR4 catalog {\sf autofit} temperatures for helium-rich stars are generally unreliable near 
or below 10\,000\,K, as their helium atmosphere models do not extend below this temperature 
\citep{eis06}.  Owing to this fact, the DC star spectroscopic and photometric data were fitted 
using the same technique and helium-rich (metal-free) models as in \citet{duf07}, yielding 
effective temperatures and photometric distances, again assuming $\log\,g=8.0$.  All 
photometric data were taken from the DR4 white dwarf catalog \citep{eis06}, with stellar 
parameters for the DZ stars from \citet{duf07} and the DC model fits performed for this work.

Spatial and kinematical data for these white dwarfs were obtained from the SDSS seventh 
data release (DR7; \citealt{aba09}).  This latest data release includes the USNO-SDSS derived 
proper motion catalog of \citet{mun04} and its recent amendments \citep{mun08}, with statistical 
errors around 3\,mas\,yr$^{-1}$.  For a handful of white dwarfs, proper motions were not available 
in the DR7 catalog; for these stars measurements were taken from the SuperCOSMOS Science 
Archive (SSA; \citealt{ham01}), the LSPM catalog \citep{lep05}, or were calculated for this work 
based on two or more positions on archival photographic plates and SDSS images separated 
by approximately 50 years.  These proper motions are listed in Table \ref{tbl1}.

\begin{table}
\begin{center}
\caption{White Dwarf Proper Motions Unavailable in SDSS DR7\label{tbl1}} 
\begin{tabular}{@{}lrrr@{}}
\hline

SDSS White Dwarf			&$\mu$			&P.A.	&Sources\\
 						&(mas\,yr$^{-1}$)	&(deg)	&\\
 
\hline
 
J000557.20$+$001833.3		&209.3		&93.5		&1,2,3\\
J020001.99$+$004018.4		&33.2		&87.7		&1\\
J020132.24$-$003932.0		&90.7		&184.1		&1\\
J080211.42$+$301256.7		&59.4		&199.8		&3\\
J082927.85$+$075911.4		&168.0		&258.7		&1\\
J083434.68$+$464130.6		&225.1		&133.2		&1,2,3\\
J084911.86$+$403649.7		&96.3		&233.8		&1\\
J093545.45$+$003750.9		&66.2		&227.0		&1\\
J094530.20$+$084624.8		&67.4		&255.7		&1\\
J113711.28$+$034324.7		&176.5		&230.3		&2,3\\
J122204.48$+$634354.5		&66.6		&251.1		&1\\
J140316.91$-$002450.0		&63.6		&96.9		&3\\
J144022.52$-$023222.2		&224.3		&271.8		&3\\
J153032.05$+$004509.0		&71.7		&249.7		&1\\

\hline
\end{tabular}
\end{center}
Sources:  (1) SSA; (2) LSPM catalog; (3) this work.

\end{table}

\section{ABUNDANCE ANALYSIS}

The SDSS white dwarfs offer a deeper snapshot of the Solar neighborhood and beyond.  
Owing to the $g>22$ AB magnitude limit of the survey, and its general avoidance of the 
Galactic plane, these white dwarfs typically represent distant stars (tens to hundreds of\,pc) 
at intermediate Galactic latitudes.  Hence the SDSS DZ stars are an excellent probe of the 
local ISM (LISM) both within and beyond the Local Bubble or Chimney \citep{wel99,wel94},
which extends roughly to 100\,pc \citep{red08}.

A number of relevant quantities were calculated from the DZ and DC white dwarf spatial 
and kinematical data:  tangential speed $v_{\rm tan}$, height above (or below) the Galactic 
mid-plane $|z|$, space velocities $UVW$ (assuming zero radial velocity), and total space 
velocity $T^2=U^2+V^2+W^2$.  Space velocities were corrected to the Local Standard of 
Rest (LSR; \citealt{deh98}), with $U$ positive toward the Galactic anticenter, $V$ positive 
in the direction of Galactic rotation, and $W$ positive toward the north Galactic pole.  In the 
absence of correction to the LSR, the assumption of zero radial velocity is equivalent to 
$T={v_{\rm tan}}$.  Tangential speeds were calculated from
\smallskip 

\begin{equation}
v_{\rm tan} \, {\rm (km\,s^{-1})} = 4.7405 \times d \, {\rm (pc)}  \times \mu \, {\rm (arcsec\,yr^{-1})} 
\label{eqn1}
\end{equation}

\smallskip\noindent
The Sun does not have zero velocity relative to the LSR; depending on viewing direction, the 
Solar motion contributes $v_{\rm tan}=10-15$\,km\,s$^{-1}$ \citep{deh98,mih81}.  Table \ref{tbl2} 
lists the calculated parameter means for the SDSS DZ and DC white dwarfs.

\begin{table}
\begin{center}
\caption{Statistical Properties of SDSS DZ and DC White Dwarfs\label{tbl2}} 
\begin{tabular}{@{}lrr@{}}
\hline

Parameter				&146 DZ Stars			&142 DC Stars\\
 				
\hline
 
$T_{\rm eff}$ (K)			&$8700\pm1330$		&$9040\pm1610$\\
$g-z$					&$-0.29\pm0.23$		&$-0.18\pm29$	\\
$d$ (pc)					&$196\pm91$			&$206\pm96$\\
$|z|$ (pc)					&$140\pm67$			&$148\pm72$\\
						&					&\\
$v_{\rm tan}$ (km\,s$^{-1}$)	&$48\pm32$			&$52\pm25$\\
$U$ (km\,s$^{-1}$)			&$2\pm35$			&$-4\pm38$\\
$V$ (km\,s$^{-1}$)			&$-12\pm28$			&$-12\pm24$\\
$W$ (km\,s$^{-1}$)			&$1\pm25$			&$-1\pm25$\\
$|W|$ (km\,s$^{-1}$)			&$18\pm17$			&$19\pm15$\\
$T$ (km\,s$^{-1}$)			&$43\pm29$			&$48\pm23$\\

\hline
\end{tabular}
\end{center}

Space velocities were calculated assuming zero radial velocity and corrected to the LSR 
(otherwise $T={v_{\rm tan}}$).

\end{table}

\subsection{A Glance at the Data}

\begin{figure}
\includegraphics[width=84mm]{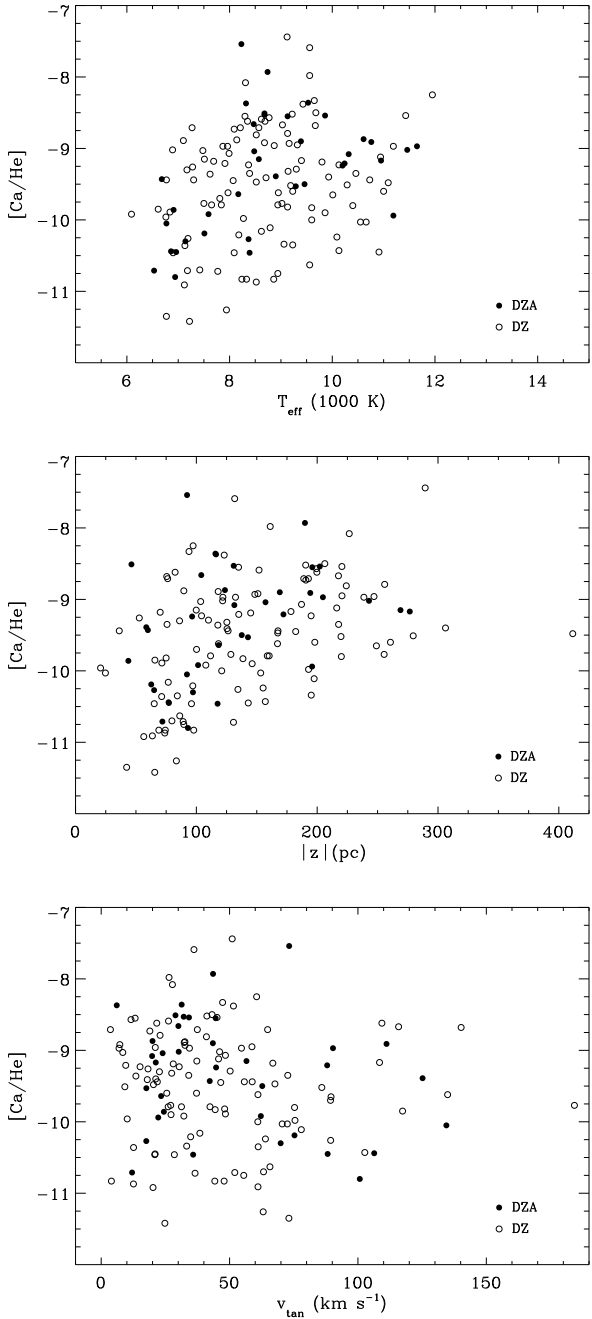}
\caption{Calcium abundances for the 146 DZ stars plotted versus effective temperature, height 
above the Galactic mid-plane, and tangential speed.  Those stars with detected hydrogen are 
plotted as filled circles, while those without are plotted with open circles.
\label{fig1}}
\end{figure}

As a first attempt to establish the relationship, if any, between the DZ stars and the ISM,
one can search for spatial or kinematical correlations with calcium abundance \citep{zuc03,
aan93}.  Figure \ref{fig1} plots the calcium abundances (throughout this paper [X/Y] = $\log
\,[n({\rm X})/n({\rm Y})]$), relative to helium, of the 146 DZ stars as a function of effective 
temperature, height above the Galactic mid-plane, and tangential speed.  A low abundance 
cutoff at higher effective temperatures and larger distances is apparent in the upper and middle 
panels; these are observational biases arising from higher atmospheric opacities and lower 
spectroscopic sensitivity, respectively \citep{duf07,koe05a}.  The classical conundrum of the 
DZ white dwarfs presents itself well in the upper panel; the majority of stars have no detected 
hydrogen, including those with the highest calcium abundances.

The middle and lower panels of Figure \ref{fig1} are noteworthy because they fail to reveal a 
physical correlation between accreted calcium and height above the Galactic disk or tangential
speed.  The Galactic gas and dust layer extends roughly 100\,pc above the Galactic mid-plane,
and yet some of most highly polluted DZ white dwarfs lie well above this region.  The lower 
panel is relevant in the context of gravitationally driven accretion, expected if the DZ stars 
obtain their heavy elements while moving through the ISM.  The lack of correlation between 
speed and the metal abundances argues, qualitatively but strongly, against fluid accretion of 
ISM.

\begin{figure}
\includegraphics[width=84mm]{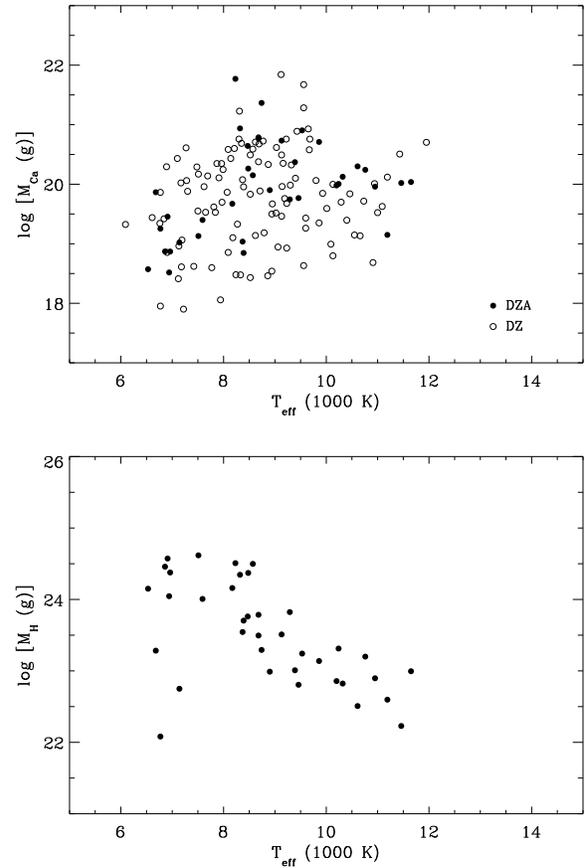}
\caption{Mass of calcium and hydrogen in the convective envelopes of the SDSS DZ stars;
envelope masses were taken from \citet{koe09}.  A typical calcium mass in the convection 
zone of these stars is $6\times10^{19}$\,g, roughly the same as that contained in the asteroid
Minerva, whose mean diameter is 146\,km and total mass is $3\times10^{21}$ g.
\label{fig2}}
\end{figure}

Figure \ref{fig2} plots the masses of calcium contained in the convection zones of these
metal-enriched stars using their abundances together with the convective envelope masses
for $\log\,g=8$ helium-rich white dwarfs \citep{koe09}.  The figure shows that the most highly 
polluted DZ stars currently harbor up to 10$^{22}$\,g of calcium in their convection zones.  
This is a truly remarkable amount of a single heavy element, a mass in calcium alone that 
is roughly equivalent to the total mass of a 200\,km diameter asteroid(!).  A more typical DZ 
star appears to contain around 10$^{20}$\,g of calcium, still a prodigious amount, and 
corresponds to a time-averaged metal accretion rate of $2\times10^8$\,g\,s$^{-1}$ 
\citep{far09a}, and a total accreted mass of metals of just under $10^{22}$\,g.

\begin{figure}
\includegraphics[width=84mm]{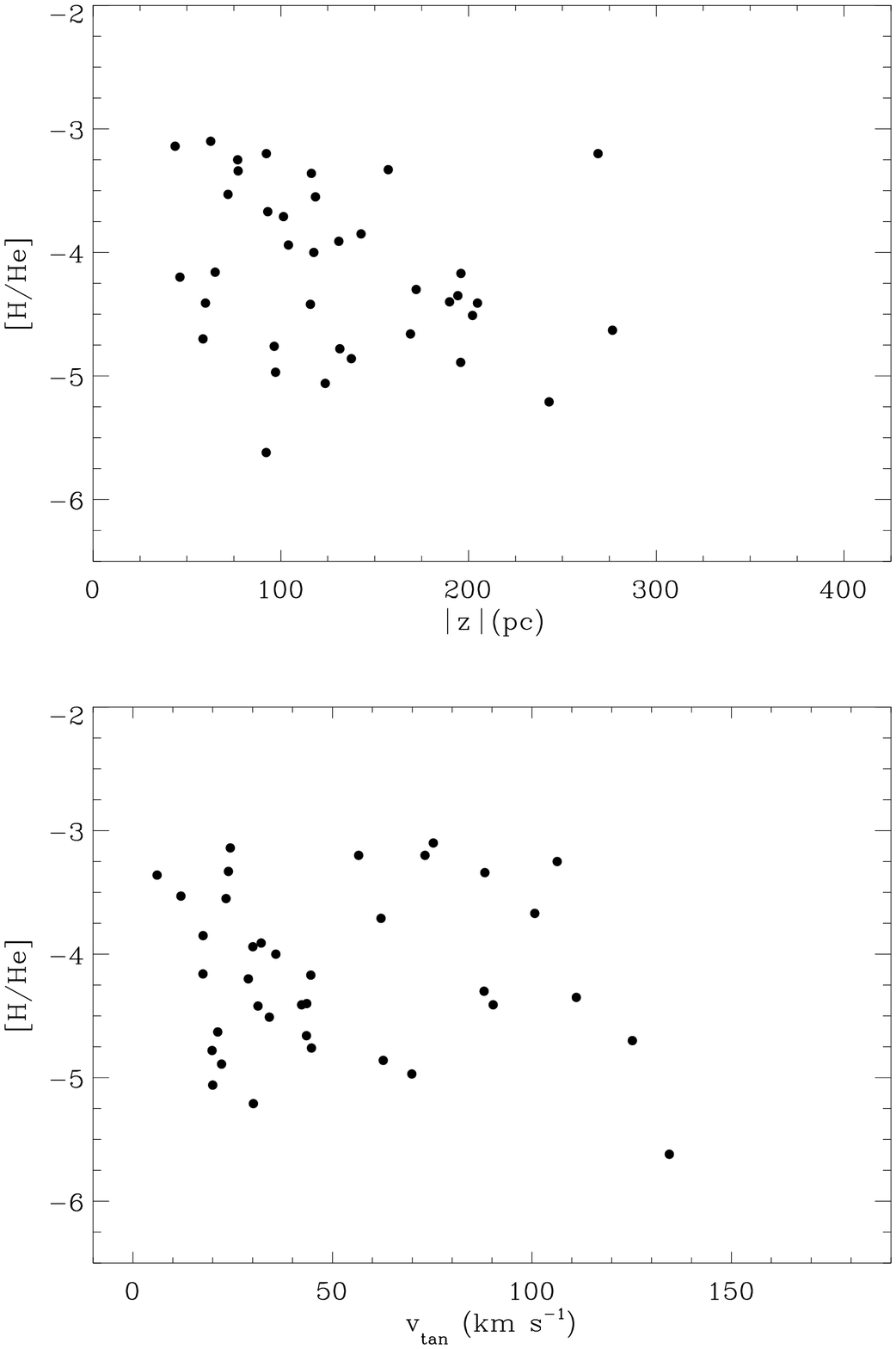}
\caption{Hydrogen abundances for the 37 DZA stars plotted versus height above the 
Galactic mid-plane, and tangential speed.
\label{fig3}}
\end{figure}

Figure \ref{fig3} plots the hydrogen abundances in 37 of the 146 DZ stars where hydrogen is
detected or inferred \citep{duf07}, versus height above the Galactic mid-plane and tangential 
speed.  No obvious pattern is seen, although there may be a higher density of DZA stars near 
the Galactic disk and perhaps also toward more modest speeds, but the former may be an 
observational bias due to diminished spectroscopic sensitivity.  In any case, one should
expect a correlation between these quantities if the hydrogen were accreted from ISM.

As discussed by \citet{dup93a}, white dwarfs accreting Solar abundance matter within the 
ISM at $3\times10^{11}$\,g\,s$^{-1}$ for 10$^6$\,yr can accumulate the heavy element mass 
fractions observed in the convective envelopes of DZ stars, such as plotted in Figure \ref{fig2}
(however, this cannot account for their hydrogen abundances, discussed in detail below). Yet 
some care must be taken not to invoke such high accretion rates without skepticism; while 
these rates are sufficient to account for the calcium data (that is the nature of the hypothesis),
the physical plausibility is of equal importance.  Below, the problem of interstellar accretion 
onto white dwarfs is reviewed in some detail and updated with old and new theoretical and
empirical considerations.

\subsection{A Closer Look at the Interstellar Accretion Hypothesis: Hydrogen}

Both \citet{koe76} and \citet{wes79} were among the first to consider the issue of interstellar
accretion onto helium-rich white dwarfs, both concluding that the atmospheres of these stars
(both type DB and DC) were {\em incompatible} with Bondi-Hoyle accretion of interstellar 
hydrogen.  That is, in most cases a single interstellar cloud encounter lasting around 10$
^5$\,yr would suffice to transform a DB (or DC) star into a DA \citep{koe76}.  Therefore, the 
existence of more than 1000 hydrogen-poor white dwarfs in the SDSS \citep{eis06} argues 
strongly against the Bondi-Hoyle type accretion of interstellar hydrogen.

Regardless of particulars relevant only to the metal-enriched white dwarf varieties, the lack 
of hydrogen in helium atmosphere white dwarfs is of great observational significance.  High 
resolution and high signal-to-noise spectroscopy reveals roughly equal numbers of DB stars
where hydrogen is either undetected or detected but deficient by a few to several orders of 
magnitude relative to helium \citep{vos07}.  The highest mass of hydrogen seen in a DB 
(metal-free) atmosphere is around $8\times10^{22}$\,g in a 12\,000\,K star, while the older 
and cooler counterparts studied here (the DZA stars) contain up to $4\times10^{24}$\,g. This 
higher mass of atmospheric hydrogen would be accreted in under 10$^6$\,yr at the interstellar 
Bondi-Hoyle rate assumed by \citet{dup93a}, and hence this cannot be the correct picture due 
to the lack of such hydrogen masses in DB stars. 

Gravitational accretion of ISM particles by a star of mass $M$ and radius $R$, in the supersonic 
regime follows the mathematical form \citep{alc80}
\smallskip

\begin{equation}
\dot M = \pi \left( R_A + R \right) R s \rho_{\infty}
\label{eqn2}
\end{equation}

\smallskip\noindent
where $R_A$ is the accretion radius, $s=\sqrt{v^2 + {c_s}^2}$ for relative stellar velocity $v$
and ambient sound velocity $c_s$, while $\rho_{\infty}$ is the unperturbed density of material 
being accreted.  The accretion radius is defined as
\smallskip

\begin{equation}
R_A \equiv \frac{2 G M}{s^2}
\label{eqn3}
\end{equation}

\smallskip\noindent
and is often referred to the as the Bondi or Bondi-Hoyle radius.  For all possible speeds
considered here, $R_A\gg R$, and the accretion rate is given by
\smallskip

\begin{equation}
\dot M = 2 \pi G M R_A R s \rho_{\infty}
\label{eqn4}
\end{equation}

\smallskip\noindent
or
\smallskip

\begin{equation}
\dot M_{\rm Edd} = \frac{2 \pi G M R \rho_{\infty}}{s} 
\label{eqn5}
\end{equation}

\smallskip\noindent
This is the Eddington rate \citep{edd26}, the accretion induced on non-interacting particles by 
the geometrical-gravitational cross section of the star as it travels through the ISM.

Bondi-Hoyle theory (i.e. including gas pressure; \citealt{edg04}) demonstrates that the effective 
cross section in Equation \ref{eqn4}, $\pi R_AR$ becomes $\pi {R_A}^2$ in the fluid dynamical 
limit \citep{bon52}, yielding an accretion rate for interacting particles
\smallskip

\begin{equation}
\dot M_{\rm BH} = \frac{4 \pi G^2 M^2 \rho_{\infty}}{s^3} 
\label{eqn6}
\end{equation}

\smallskip\noindent
This mass infall rate represents the maximal, idealized case where the mean free path of the
particles is such that collisions are important and transverse momentum is effectively destroyed 
downstream from the star.  It is also physically unrealistic in perhaps all situations excepting an
ionized plasma, as it assumes no net angular momentum between the accreting star and its
surrounding medium \citep{koe76}, and it certainly does not apply to neutral atoms or large 
particles \citep{alc80}.  The ratio of the Bondi-Hoyle to Eddington accretion rates is $v^2R/2GM$ 
for $v\gg c$, or around 10$^4$ for typical white dwarf sizes and speeds \citep{koe76}.  Table 
\ref{tbl3} lists typical densities and other relevant parameters for four fundamental types of ISM: 
molecular clouds, diffuse clouds, warm ionized, and hot ionized.  Listed also are the expected 
high-end mass infall rates for white dwarfs moving through these regions following either 
Bondi-Hoyle (fluid) or Eddington (geometric) type accretion.

\begin{table*} 
\begin{minipage}{100mm}
\begin{center}

\caption{Representative High ISM Accretion Rates for White Dwarfs\label{tbl3}} 
\begin{tabular}{@{}llcccll@{}}
\hline

ISM Type			&Gas		&$\rho_{\infty}$		&$T$		&$c_s$		&$\dot M_{\rm BH}$		&$\dot M_{\rm Edd}$\\
				&			&(cm$^{-3}$)		&(K)			&(km\,s$^{-1}$)		&(g\,s$^{-1}$)			&(g\,s$^{-1}$)\\ 				
 
\hline
 
Molecular Cloud 	&H$_2$		&1000			&20			&0.3				&$10^{12}$			&$10^8$\\
Diffuse Cloud 		&H			&100			&80			&1				&$10^{11}$			&$10^7$\\
Warm Ionized 		&H$^+$		&1				&8000		&1				&$10^{9}$			&$10^5$\\
Hot Ionized 		&H$^+$		&0.01			&10$^6$		&100			&$10^7$				&$10^3$\\

\hline
\end{tabular}
\end{center}

Accretion rates listed are order of magnitude estimates calculated for the listed parameters 
and $v=50$\,km\,s$^{-1}$, a speed typical of both the DZ and DC samples.

\end{minipage}
\end{table*}

The two-phase accretion-diffusion scenario as laid out by \citep{dup93a}, invokes 10$^6$\,yr
within a cloud (accretion), and $5\times10^7$\,yr between clouds (diffusion).  For disk stars
with Galactic orbits similar to the Sun, these timescales imply about five cloud encounters per
240\,Myr orbit.  The motions of interest are the relative motions between stars and the ISM, the 
latter which should be moving within the relatively low velocity spiral arms of the Galactic disk 
\citep{jah97}.  

If correct, this implies a typical helium-rich white dwarf moving at 50\,km\,s$^{-1}$ relative to 
the LSR and (presumably) the ISM travels, on average, 240\,pc within clouds during a single 
Galactic rotation.  This timescale corresponds to the cooling age for a $\log\,g=8.0$, 14\,500\,K 
helium-rich white dwarf, and hence such a star should, according to this scenario, obtain up to 
$5\times10^{22}$\,g of hydrogen via geometric accretion, or [H/He] $=-4.4$.  This value is 
commensurate with the highest hydrogen abundances observed in DB stars, and well 
above the lower limit of detectability \citep{vos07}.

This comparison implies 1) the densities and corresponding Eddington accretion rates in Table 
\ref{tbl3} are too high, or 2) cloud encounters last less than 10$^6$\,yr or are less frequent than 
once per $5\times10^7$\,yr, but is otherwise consistent with the geometric capture of hydrogen 
within the ISM, and inconsistent with fluid accretion.  Ignoring the fact that all hydrogen will be 
ionized within the Bondi-Hoyle radius of the white dwarf, and hence should accrete as a plasma
at the fluid rate \citep{alc80}, it is clear that fluid accretion of hydrogen does not occur in helium 
atmosphere white dwarfs, if and when passing through dense ISM.

\begin{figure}
\includegraphics[width=84mm]{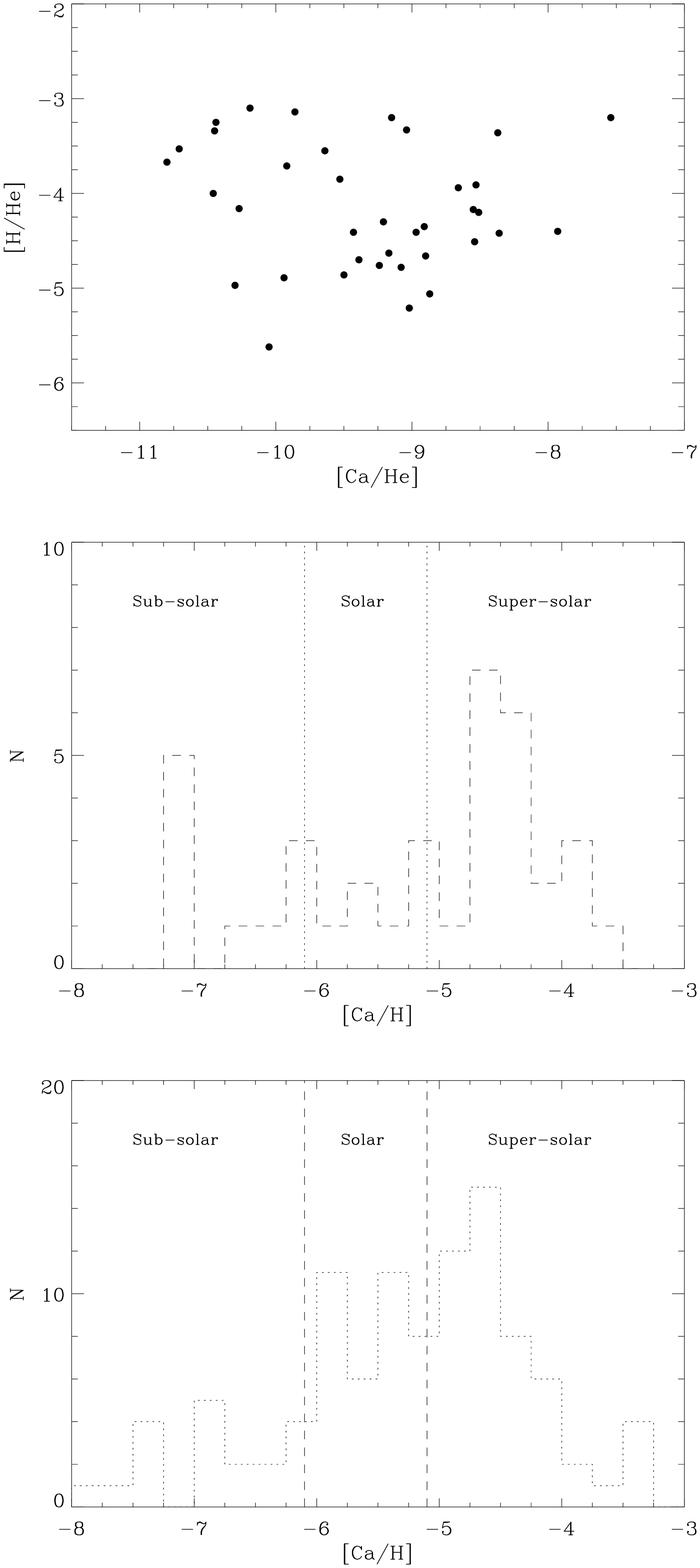}
\caption{The upper panel plots the hydrogen versus calcium abundances for the 37 DZA
stars, while the middle panel plots a histogram of their [Ca/H] ratios.  The lower panel plots 
[Ca/H] ratio {\em lower limits} for the 109 DZ white dwarfs without detected hydrogen.  The 
Solar abundance is [Ca/H] $=-5.6$.
\label{fig4}}
\end{figure}

\subsection{Super-solar Calcium to Hydrogen Abundances}

The average DZ star in the sample has an effective temperature of 9000\,K and a cooling age 
around 860 Myr \citep{fon01}.  A hydrogen abundance of [H/He] $=-3.8$ was detectable in all 
146 DZ stars \citep{duf07}, equivalent to a bit less than 10$^{24}$\,g of hydrogen at the average 
DZ temperature.  In order to avoid such a hydrogen mass within its outer envelope, a helium-rich 
white dwarf must eschew 1) fluid accretion in any region similar to the relatively low density 
LISM for 99.9\% of its cooling age, and 2) geometrical accretion in any molecular cloud, for 
around 90\% of its cooling age.  Only 12 of the DZ stars or 8\% of the sample have hydrogen 
abundances [H/He] $\geq-3.8$, implying such avoidance has occurred for the remaining 134 
DZ(A) stars.  If the photospheric calcium originates in the ISM, then a process that accretes 
metal-rich dust grains while eschewing hydrogen (and other gaseous volatiles) is necessary 
to account for the DZ white dwarfs with little or no hydrogen.

Historically, the super-solar [Ca/H] abundances of DZ white dwarfs were qualitatively accounted
for by preferential accretion of dust grains \citep{dup93b}.  A hypothesis to permit the incursion 
of dust grains yet repulse light gas is the propeller mechanism, whereby charged ion species 
are repelled at the Alfv\'en radius \citep{wes82}.  This model fails to explain a number of very 
hydrogen-deficient DZ-type stars, including GD 40 with [H/He] $=-6.0$, a white dwarf for which 
\citet{fri04} failed to find evidence of magnetism sufficient to drive a propeller.  Rather, in the same 
study the DZA star LHS 235, with 400 times more hydrogen than GD 40 was found to have a 7 kG 
magnetic field -- essentially {\em the exact opposite} of what might be expected if hydrogen was 
screened efficiently by a propeller.  Perhaps the coup de gr\^ace for the propeller is the magnetic 
DZA star G165-7; with a 650 kG field, [Ca/He] $=-8.1$, and [H/He] $=-3.0$ \citep{duf06}, it has 
everything a `fan' could hope for.  But should this star have so much hydrogen?  According to 
theory, its propeller failed around 660 Myr ago as a 7500\,K star \citep{wes82}, and could have 
amassed at most $2\times10^{24}$\,g in 660 Myr of {\em continuous} accretion at the high 
Eddington rate, corresponding to [H/He] $=-4.0$.  Clearly, this star was never a propeller.  This 
mechanism is unsupported by observations \citep{fri04}, and is fundamentally at odds with the 
existence DBAZ and DZA stars \citep{vos07}.

The upper panel of Figure \ref{fig4} plots the hydrogen versus calcium abundances of all 
37 DZA stars.  The lack of any correlation between the two elements suggests either 1) 
disparate origins, or 2) a common origin with intrinsic variations in [Ca/H].  Neither of these 
two possibilities is consistent with ISM accretion.  The middle and lower panels of Figure 
\ref{fig4} display the inferred [Ca/H] abundances and lower limits for the 37 DZA and 109 DZ 
stars.  In both cases, the vast majority of these values are super-solar by at least an order of 
magnitude (see also Figure 12 of \citealt{duf07}).

\subsection{A Closer Look at the Interstellar Accretion Hypothesis: Metals}

Interstellar gas -- both cold and warm -- is metal-poor; heavier elements tend to be locked up 
in grains, leaving only light and volatile elements \citep{ven90}.  This fact was not considered 
by \citet{dup93a}, whose accretion-diffusion scenario explicitly assumes 1) accretion based 
on fluid rates for hydrogen also apply to heavy elements, and 2) elements are accreted in 
Solar proportions.  Above it was shown that fluid accretion rates for ISM accretion of hydrogen 
over predict the observed abundances in DB stars by a few to several orders of magnitude.  
But more importantly for heavy element accretion is the fact that {\em interstellar dust grains 
should not accrete at the fluid rate}.

Solid particles will follow trajectories independent of surrounding gas whether neutral, or 
charged and electrically or magnetically coupled to the gas \citep{alc80}.  However, dust grains 
approaching the stellar surface will be separated into their constituent elements via sublimation 
that proceeds at slightly different rates according to the particular chemical species, resulting 
in significant ingress prior to becoming gaseous.  The gradual evaporation of approaching 
interstellar dust has two major consequences.  First, gaseous heavy elements cannot accrete 
at the fluid rate, as the effective gravitational cross section is significantly reduced by the close 
approach necessary for their evaporation out of grains.  Second, the process of sublimation 
will spatially and temporally separate heavy elements according to their specific evaporation 
temperatures, resulting in preferentially higher accretion rates for volatile species such as ice
mantles, as opposed to metallic atoms or refractory material.  This latter consequence is 
interesting because it predicts a relatively high volatile to refractory element accretion mixture; 
in stark contrast both to the observed abundances and to those expected from the accretion of 
circumstellar material \citep{sio90a}.

The radius at which dust grains are evaporated, $R_{\rm ev}$, and become part of the 
ambient, interacting gas is substituted for $R$ in Equation \ref{eqn2} which then becomes 
\citep{alc80}
\smallskip

\begin{equation}
\dot M_{\rm gr} =  \pi \left( R_A + R_{\rm ev} \right) R_{\rm ev} s \rho_{\infty}
\label{eqn7}
\end{equation}

\smallskip\noindent
If the grains absorb and radiate energy as blackbodies, they will evaporate at temperature
$T_{\rm ev}$ at a distance from the star given by \citep{che01}
\smallskip

\begin{equation}
R_{\rm ev} \approx \frac{R}{2} \left( \frac{T_{\rm eff}}{T_{\rm ev}} \right)^2
\label{eqn8}
\end{equation}

\smallskip\noindent
Taking a typical DZ effective temperature of 9000\,K from Table \ref{tbl3} and $T_{\rm ev}$ 
between 1500 and 2000\,K, the resulting evaporation radius is between 10 and 18 white 
dwarf radii (or 0.13 to 0.23\,$R_{\odot}$).  This is a few hundred times smaller than a typical 
Bondi-Hoyle radius, around 0.5\,AU for a white dwarf, and hence heavy elements will accrete 
at a rate much closer to the Eddington rate.  In this situation $R_A\gg R_{\rm ev}$, and the 
resulting mass infall becomes
\smallskip

\begin{equation}
\dot M_{\rm gr} = \frac{\pi G M  R \rho_{\infty}}{s} \left( \frac{T_{\rm eff}}{T_{\rm ev}} \right)^2
\label{eqn9}
\end{equation}

\smallskip\noindent
All else being equal, this rate should be around 10 to 18 times the Eddington rate for 
refractory-rich dust grains, such as interstellar silicates.  As pointed out more than thirty
years ago by \citet{koe76}, the above analysis ignores any net angular momentum between 
the star and the surrounding ISM, which will work to reduce the effectiveness of accretion
excepting where material is fully ionized, a situation unlikely to be realized for gaseous
heavy elements.  The physical considerations discussed in this section have been, for the 
most part, overlooked in several major studies of DZ stars, and their absence has biased 
outcomes favoring the interstellar origin of metals in cool white dwarfs in general 
\citep{koe06,wol02,fri00,dup93b}.

Applying Equation \ref{eqn8} to the highest density, Solar abundance ISM (i.e. a dust to 
gas ratio of 1:100, [Ca/H] $=-5.6$, $\rho_{\infty}=1000$\,g\,cm$^3$), a white dwarf can accrete 
heavy elements from the ISM at up to $2\times10^7$\,g\,s$^{-1}$.  Over 10$^6$\,yr of sustained 
accretion at this rate, a typical white dwarf would accumulate $7\times10^{18}$\,g of calcium in 
its convective envelope; longer timescales are irrelevant because diffusion begins to compete 
with accretion.  This quantity of calcium is able to account for a small minority of the observed 
envelope masses plotted in the upper panel of Figure \ref{fig2}; the average calcium mass is 10 
times higher, with a few values 3 orders of magnitude greater.  This calculation assumes 1) 
the highest accretion rate for dust grains within dense ISM regions, and 2) stars that currently 
retain their peak calcium abundances to within an order of magnitude.  Hence, the calculation
should {\em over predict} the amount of calcium polluting the stellar convection zones by one 
to a few orders of magnitude.  Therefore, if this analysis is valid to within an order of magnitude, 
it is all but certain that the photospheric calcium in DZ stars cannot have interstellar origins.

The DAZ white dwarfs make this point even more striking.  For these polluted, hydrogen
atmosphere stars with thin convection zones and short metal diffusion times relative to the 
DZ stars, \citet{koe06} infer continuous and ongoing mass accretion rates based on a highly
probable, current steady state balance between accretion and diffusion.  The inferred DAZ
accretion rates for Solar abundance matter, up to $2\times10^{11}$\,g\,s$^{-1}$, are quite 
similar to the high rate assumed by \citet{dup93a} for the DZ stars.  Yet no correlation is 
found between the implied DAZ accretion rates and the total space velocity of the stars 
as expected in the case of fluid accretion \citep{koe06}.  Furthermore, there are no known 
regions of enhanced LISM density in the vicinity of the DAZ stars \citep{kil07,koe06,zuc03}, 
and Table \ref{tbl3} reveals that accretion rates of $10^{11}$\,g\,s$^{-1}$ are only even 
theoretically possible within relatively dense clouds, which are not seen.  Lastly, there is 
firm evidence that the nearby DAZ G29-38 does not accrete matter at the fluid rate, but 
rather has an upper limit based on X-ray observations that is 25 times lower \citep{jur09b}.

There is no physical reason why helium and hydrogen atmosphere white dwarfs should 
become polluted via separate mechanisms; in fact, Occam's razor argues against a bimodal 
theory.  The fact that some examples of both varieties of metal-contaminated stars are found 
with closely orbiting dust is testament to this fact \citep{far09a}.  In this regard, the only real 
difference between the DZ and DAZ stars is that the latter group is currently accreting from 
its immediate surroundings, while the former group carries their metallic scars long after a 
pollution event has ended \citep{koe09}.

\section{SPATIAL - KINEMATICAL ANALYSIS}

\subsection{Polluted White Dwarfs in the Local Bubble}

While \citet{koe06} obtain capture rates up to $10^{11}$\,g\,s$^{-1}$ in the warm, ionized 
LISM within the Local Bubble which assumes fluid accretion is valid for heavy elements, there 
are still serious problems with this picture when applied to the DZ stars.  The individually identified 
parcels of LISM span at most a few pc \citep{red08}, corresponding to several 10$^4$\,yr of 
traversal by a typical DZ white dwarf.  Such an encounter in a typical region of LISM with 
$\rho_{\infty}=0.1$\,cm$^{-3}$ \citep{red00} leads to $\dot M_{\rm BH}$ rates of $10^8$\,g\,s$
^{-1}$, and would provide only 10$^{16}$\,g of calcium pollution if accreting Solar abundance 
matter.  While such modest an amount of metal can effectively pollute the relatively thin outer 
layers of DAZ stars, this paltry calcium mass requires 10\,000 LISM encounters to account for 
the average calcium mass in the DZ star envelopes.  Of course, this presumes fluid accretion
is viable for heavy elements, which has been shown above to be physically implausible, thus 
making legion fluid infall episodes unlikely in the extreme.  Therefore, it is safe to conclude 
that within the Local Bubble, ISM accretion is inconsistent with the phenomenon of DZ stars, 
which on average travel only 50\,pc during a 10$^6$\,yr metal sinking timescale.

\subsection{ISM Accretion Above the Galactic Disk?}

Prior to the SDSS, there were few (if any) cool ($T_{\rm eff}\leq12\,000$\,K) white dwarfs 
known beyond 200\,pc \citep{mcc99}.  Generally a hindrance for both magnitude limited as 
well as proper motion surveys, such distant cool degenerates would have $V>18$\,mag, while 
100\,km s$^{-1}$ tangential speeds only yield $\mu\la100$\,mas\,yr$^{-1}$.  The DZ sample of 
\citet{aan93} was limited to within 50\,pc of the Sun, but the SDSS DZ sample stretches out to 
300\,pc {\em above the disk of the Milky Way} (see Figure \ref{fig1}).  While the single DZ outlier 
at $|z|>400$\,pc is likely to be a higher than average mass white dwarf (i.e. its photometric 
distance is overestimated), the sample as a whole contains 96 (66\%) and 26 (18\%) stars 
situated more than 100 and 200\,pc above the plane of the Galaxy.  This fact alone perhaps 
provides an insurmountable challenge to the ISM accretion hypothesis.

The scale height of molecular gas as measured by CO is $75\pm25$\,pc \citep{dam85}, while
that of neutral, atomic hydrogen is roughly 135\,pc with substantial irregularities \citep{loc86}. 
Taking 100\,pc as a fiducial scale height for interstellar gas and dust, and assuming the entire 
sample of DZ stars obtained their metals from this medium requires the stars: 1) are currently 
traveling through or away from the Galactic mid-plane, and 2) have not traveled significantly 
further than around 18\,pc above the gas and dust layer, on average.  This distance constraint 
corresponds to a star traveling at the sample mean $|W|=18$\,km\,s$^{-1}$ speed (see 
Table \ref{tbl2}) over a typical Myr diffusion timescale.  

\begin{figure}
\includegraphics[width=84mm]{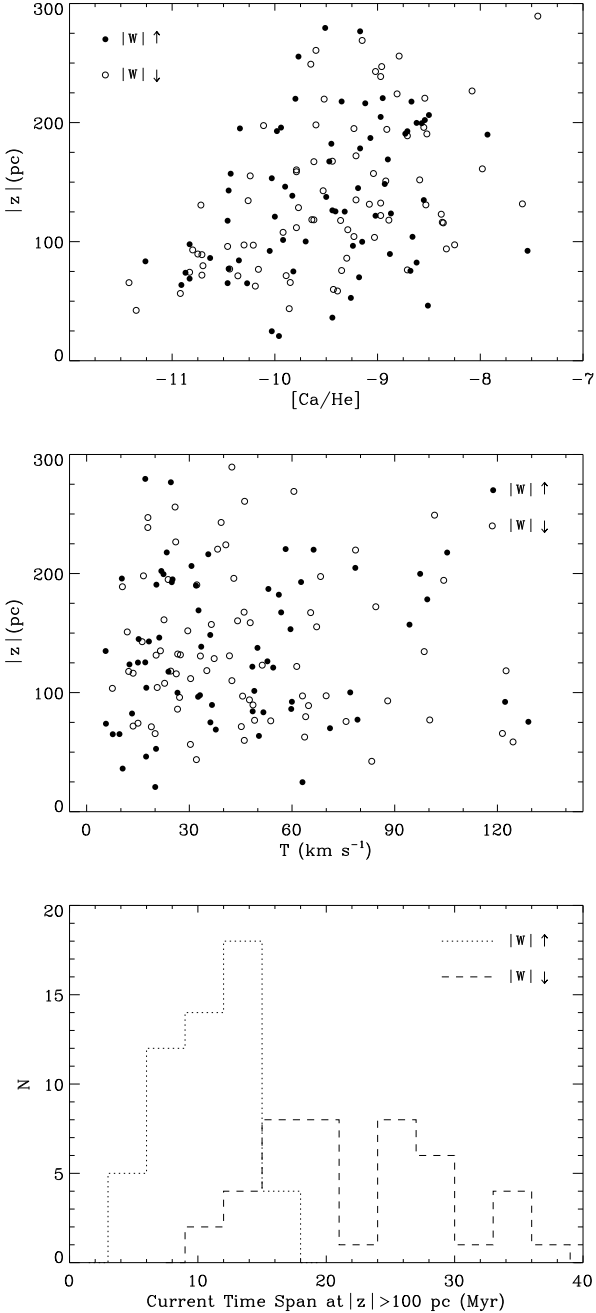}
\caption{The upper panel reveals 67 DZ stars moving back toward the Galactic mid-plane,
many among the most highly metal-polluted stars in the sample.  The middle panel shows
the total space velocity as a function of height above the Galactic disk, with many stars 
exhibiting high values along both axes.  The lower panel is a histogram of Myr spent at 
$|z|>100$\,pc for the 96 DZ stars currently located in this region.
\label{fig5}}
\end{figure}

However, these requirements do not match the observations.  First, more than half the sample 
(81 stars or 58\%) currently sit more than 120\,pc above the disk of the Galaxy.  Second, and 
rather damningly, {\em nearly half of the sample is moving back into the disk rather than away}.
The upper	 panel of Figure \ref{fig5} reveals 67 polluted white dwarfs that are currently moving
back toward the Galactic mid-plane.  Many of these stars are among the most highly polluted 
in the sample, and many have $|z|\ga100$ pc.  The middle panel of the same figure shows the
height above the disk as a function of total space velocity; surprisingly some of the fastest stars
are also high above the plane.  These results strongly conflict with expectations from ISM 
accretion models.

In order to estimate how much time high $|z|$ DZ stars spend above the spiral arms and 
interwoven disk material, one can use the Solar motion as a first estimate.  Depending on the 
Galactic model potential, and whether it includes dark disk matter, the Sun oscillates about 
$z=0$ every 60 to 90\,Myr \citep{mat95,bac85,mih81}, the former and latter values derived 
using models containing substantial and no dark disk matter, respectively.  \citet{kui89} provide 
$W_{\rm max}$ and $z_{\rm max}$ data for 100\,pc $\leq z\leq2000$\,pc based on a sample 
of K dwarfs toward the South Galactic Pole and a Galactic model which contains no dark disk 
component.  Using a simple harmonic oscillator model, a period can be computed from these 
maximum heights and speeds for stars within several hundred pc of the disk.  The period of 
a simple oscillator is
\smallskip

\begin{equation}
P_z = 2 \pi \sqrt{ \frac{K_z}{m}}
\label{eqn10}
\end{equation}

\smallskip\noindent
where
\smallskip

\begin{equation}
K_z = m \left( \frac{W_{\rm max}}{z_{\rm max}} \right)^2
\label{eqn11}
\end{equation}

\smallskip\noindent
In reality, the period is a function of $z$ due to the disk component of the Galaxy, which acts
as a linear term in the vertical gravitational potential, while the spherical component (including
the inferred dark halo) acts as a quadratic term, equivalent by itself to a harmonic potential 
\citep{bac92}.  Using the above equations and the data of \citet{kui89} for stars within 500\,pc
of the plane, the mean period is $P_z=80\pm10$\,Myr and compares favorably to rigorously 
derived values for the Sun.  Taking $P_z=80$\,Myr, one can then use the same simple model 
to calculate the length of time a DZ star with current velocity $W$ and height $z$ has spent
outside the 100\,pc Galactic gas and dust layer.

The lower panel of Figure \ref{fig5} plots a histogram of the outward- and inward-	bound DZ
white dwarfs which are currently at least 100\,pc above the Galactic mid-plane.  All 96 of these
stars have spent more than 3 Myr in a region of space where the accretion of interstellar matter
can be likely ruled out, 91 of these metal-polluted stars have been out of the Galactic disk for 
over 6 Myr.  As can be seen by reviewing Figure \ref{fig1}, these high $|z|$ DZ stars are not
anomalous in terms of their abundances, but rather mundane and truly representative of the
sample at large.  If anything these white dwarfs have higher abundances than the lower $|z|$ 
stars (but this is likely a bias resulting from spectroscopic calcium detection sensitivity).

If the majority of these DZ white dwarfs have traveled away from where they were polluted for 
a few to several diffusion timescales, the extant photospheric metals represent only a fraction 
of their peak abundances.  This picture dramatically exacerbates the original problem of the
metal origins in the following manner.  Instead of accounting for $10^{22\pm2}$\,g of total 
accreted metals, a viable theory must now account for nearly $10^{24\pm2}$\,g -- Ceres 
and Moon masses of heavy elements -- for a sample which has traveled for 4 to 5 diffusion 
timescales in the absence of accretion.  This possibility would greatly strengthen the need 
for efficient delivery of planetesimal-size masses of heavy elements, and combined with the 
hydrogen deficiency, is more rather than less challenging to explain with interstellar accretion.  
Hence this possibility is not considered further.

\subsection{ISM Accretion At High Speed?}

Throughout the preceding sections, a nominal 50\,km\,s$^{-1}$ tangential speed was used
for relevant calculations; this value being somewhat friendly to ISM accretion models and also
the mean among the 146 DZ stars in the sample.  However, there are 13 stars with $v_{\rm tan}
\geq100$\,km\,s$^{-1}$ (and total space velocities, $T>100$\,km\,s$^{-1}$ since their radial 
velocities are unknown and assumed to be zero), which would imply a decrease in the Table
\ref{tbl3} fluid accretion rates by an order of magnitude.  There appears to be no correlation 
between these speeds and the accreted calcium abundances in the DZ stars (see Figure 
\ref{fig1}), and hence there seems little or no chance that these stars have accreted in this 
manner.  \citet{koe06} finds a similar lack of correlation among the known DAZ stars; a more 
striking statement given their necessary ongoing accretion.

While the dependence of geometrical accretion on velocity is milder by comparison, for ISM
accretion scenarios it is reasonable to expect a mild correlation should exist between accreted 
mass and velocity, yet no such correlation is seen in Figures \ref{fig1} or \ref{fig3}.  In the middle
panel of Figure \ref{fig5}, one would expect to find milder velocities necessary to capture more 
ISM material as stars move away from the Galactic gas and dust layer and into regions where
the ISM is tenuous at best, yet this pattern is not observed.

\section{The SDSS DZ and DC White Dwarfs Compared}

If the ISM accretion hypothesis is correct, then the only difference between a DZ and a DC 
star is its recent (within a few Myr) spatial and kinematical history.  That is, during that time 
a DZ star has recently exited a relatively {\em dense} patch of ISM, while a DC star has not 
(the continuous, warm-phase ISM accretion scenario forwarded by \citet{koe06} does not 
work for the DZ stars, as demonstrated above).  Is this the correct picture? 

\begin{figure*}
\includegraphics[width=168mm]{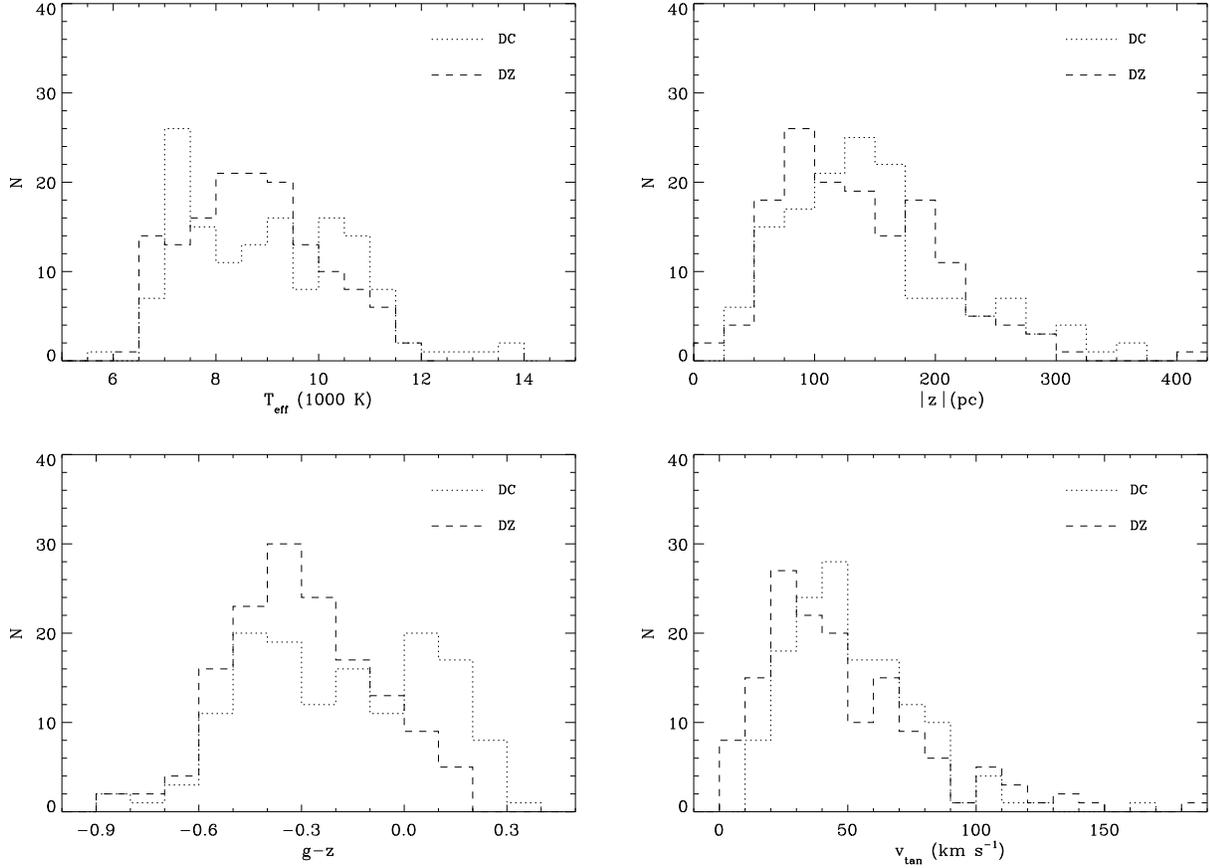}
\caption{Effective temperature, $g-z$ color, Galactic height, and tangential speed histograms for 
the DZ and DC white dwarfs.  For the most part, these two spectral classes appear to represent 
the same population of helium-rich white dwarfs.
\label{fig6}}
\end{figure*}

There are a similar number of DZ and DC white dwarfs in the SDSS and a comparison of their 
model effective temperatures, $g-z$ colors (the longest baseline possible which is unaffected 
by calcium H and K absorption), Galactic heights, and tangential speeds is shown in Figure 
\ref{fig6}.  There appear to be a few modest differences between the sample distributions; a 
small group of relatively lower temperatures, and a flatter color distribution are found among 
the DC stars, while the $|z|$ and $v_{\rm tan}$ distributions are fairly similar. Table \ref{tbl2} 
demonstrates the two spectral classes have mean parameters which agree rather well 
despite the relatively large standard deviations; their average effective temperatures, 
distances, Galactic heights, and tangential speeds all agree to within 4\% to 8\%.

\begin{figure}
\includegraphics[width=84mm]{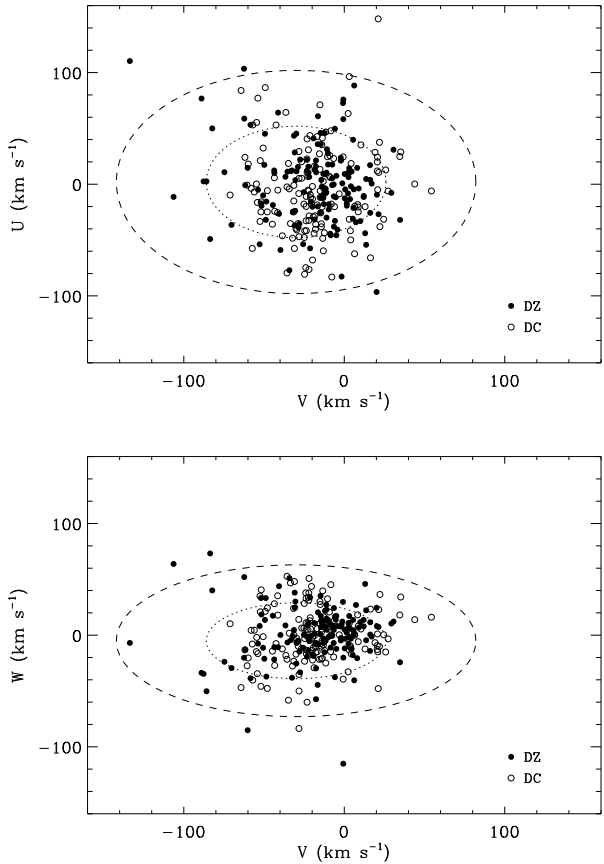}
\caption{$UVW$ space velocities for the DZ and DC white dwarfs calculated assuming 
$v_{\rm rad}$=0.  The ellipses shown in the plots are the 1 and $2\sigma$ contours for old, 
metal-poor disk stars from \citet{bee00}.
\label{fig7}}
\end{figure}

Figure \ref{fig7} plots the $UVW$ space velocities of the DZ and DC samples (assuming
$v_{\rm rad}=0$), supporting the idea they are both basically thin disk populations, each with 
several possible thick disk members.   Table \ref{tbl2} shows the two groups share the same 
mean space velocities and velocity dispersions, the latter being the most important indicator
of kinematical age \citep{jah97,mih81}.  Together these analyses suggest DZ and DC white 
dwarfs belong to the same population of helium atmosphere, thin (and a few thick) disk stars. 

\subsection{Convergent and Divergent Pairs of DZ and DC White Dwarfs}

\begin{figure}
\includegraphics[width=84mm]{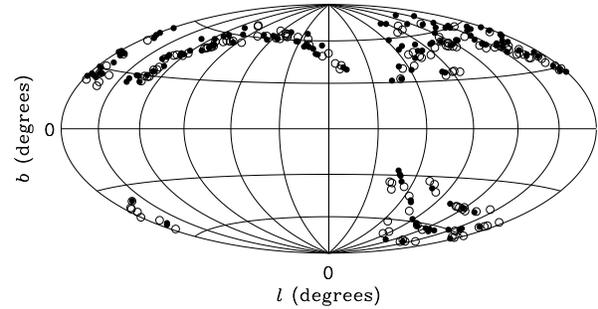}
\caption{Projection of the DZ and DC Galactic coordinates.
\label{fig8}}
\end{figure}

Figure \ref{fig8} projects the Galactic longitude and latitude of all the DZ and DC stars onto 
the plane of the sky.  While it is true that the SDSS observes only certain areas of the sky, and
hence one can a priori expect these two groups to occupy overlapping regions, the degree of
overlap between the DZ and DC white dwarfs is total; there does not exist any region where 
DZ stars are found in higher concentrations than their DC counterparts.  However, this type of
projection is very dense on paper while the actual sky is rather vast, and it necessarily ignores
the third dimension, namely the distance from the Sun.  Therefore a pertinent question to ask
is whether any of these stars are pc-scale neighbors.  A search was conducted around each
DZ and DC star for other sample stars within a $r=10$\,pc radius, and the results are listed in
Table \ref{tbl4}.  

All together 18 pairs of stars were identified in this manner, three of which lie together within 
a 5\,pc region of space.  There are nine DZ-DC, 4 DZ-DZ, and 5 DC-DC pairs; exactly half 
the pairs are of mixed spectral type.  Of the DZ-DC pairs, there are three where the total space 
velocities agree to within 12\,km\,s$^{-1}$, pairs \#1, 6, and 10.  In the latter case, each space
velocity component by itself agrees within 3\,km\,s$^{-1}$(!).  These three pairs of stars have
traveled the last 10$^6$\,yr in adjacent regions of interstellar space, yet each component star has 
evolved distinctly from its partner -- one metal-polluted and the other not.  Furthermore, there
are two DZ-DZ pairs whose total space velocities differ by more than 100\,km\,s$^{-1}$, thus
implying their current spatial convergence is in contrast to their divergent interstellar history
over the last 10$^6$\,yr.  Therefore, these two DZ-DZ pairs obtained their metals while traversing 
distinct rather than adjacent regions of the Galaxy.  

Both the DZ-DC pairs with convergent spatial histories, as well as the DZ-DZ pairs with 
divergent orbital pasts are strong counterexamples to expectations for accretion within dense 
interstellar clouds of sizes several pc or larger.  This is not surprising in the sense that such 
clouds do not exist within the Local Bubble, but again necessitates the existence of plentiful,
smaller ISM patches to account for 1) many of the pairs in Table \ref{tbl4}, and 2) the DAZ 
spectral class \citep{koe06}.  As discussed above, the scenario of small (undetectable) cloud 
encounters may work for the DAZ stars, but not for the DZ class; it requires thousands of such
encounters to accumulate the large mass of metals currently residing in their convection zones.

\begin{table*} 
\begin{minipage}{145mm}
\begin{center}

\caption{Pairs within 10 pc of one another\label{tbl4}} 
\begin{tabular}{@{}lcccccrrrr@{}}
\hline

Pair		&Star$_1$				&Star$_2$		&SpT$_1$&SpT$_2$	&$\Delta r$&$\Delta U$	&$\Delta V$	&$\Delta W$\\	
		&						&				&		&			&(pc)	&(km\,s$^{-1}$)	&(km\,s$^{-1}$)	&(km\,s$^{-1}$)\\	
 
 \hline
 
1	&J020132.24$-$003932.0		&J015604.19$-$010029.3	&DZ		&DC		&3.4		&6.1			&4.3       		&1.8\\
2	&J084828.00$+$521422.5		&J083736.58$+$542758.3	&DZ		&DC       	&5.4      	&49.3      		&57.6      		&42.9\\
3	&J084828.00$+$521422.5		&J085133.95$+$542601.2	&DZ		&DC       	&5.2      	&20.5      		&14.2      		&22.1\\
4	&J084828.00$+$521422.5		&J090729.10$+$513805.6     	&DZ		&DC       	&9.2      	&19.6       		&7.1      		&20.6\\
5	&J093423.17$+$082225.3		&J093819.84$+$071151.6     	&DZ		&DC       	&3.1      	&7.8      		&21.8      		&13.1\\
6	&J104915.06$-$000706.2		&J105632.20$-$001041.3     	&DZ		&DC       	&8.9       	&9.3       		&7.3       		&2.4\\
7	&J123455.96$-$033047.1		&J124006.36$-$003700.9     	&DZ		&DZ       	&7.6       	&2.1      		&21.3      		&12.7\\
8	&J124006.36$-$003700.9		&J124333.86$+$031737.1     	&DZ		&DC       	&8.4      	&35.3       		&0.6       		&7.4\\
9	&J131336.96$+$573800.5		&J131750.20$+$600532.9     	&DZ		&DC       	&6.7       	&9.1     		&115.1    		&55.6\\
10	&J135137.07$+$613607.0		&J135137.07$+$613607.0     	&DZ		&DC       	&6.0       	&2.4       		&1.5       		&1.6\\
11	&J144022.52$-$023222.2		&J144516.24$-$020849.6     	&DZ		&DZ       	&6.4      	&69.6      		&55.8      		&47.2\\
12	&J210733.93$-$005557.7		&J212424.69$-$011452.5     	&DZ		&DZ       	&9.2      	&80.2      		&22.2      		&57.6\\
13	&J223222.32$+$010920.7		&J223841.05$+$010150.3     	&DZ		&DZ       	&3.8       	&4.3      		&32.1      		&28.7\\
14	&J021546.95$-$073226.8		&J022147.29$-$084236.1     	&DC		&DC      	&6.2       	&7.1     		&100.0		&9.5\\
15	&J033939.67$+$001539.8		&J034401.34$-$001221.0     	&DC		&DC       	&8.6      	&23.9      		&22.1      		&29.3\\
16	&J083736.58$+$542758.3		&J085133.95$+$542601.2     	&DC		&DC       	&5.6      	&28.8      		&43.4      		&20.9\\
17	&J104718.30$+$000718.3		&J111007.61$+$011041.4     	&DC		&DC       	&9.6     	&116.8    		&17.0       		&8.3\\
18	&J130526.15$+$001250.8		&J131056.92$-$002451.9     	&DC		&DC       	&8.1    	&124.7		&35.0      		&31.1\\

\hline
\end{tabular}
\end{center}

\end{minipage}
\end{table*}

\subsection{The Frequency of Metal-Free Counterparts to DZA Stars}

A white dwarf accreting Solar ratio abundances at the high Eddington rate for 10$^6$\,yr 
will obtain about $3\times10^{21}$\,g of hydrogen, and it is tempting to imagine the hydrogen 
masses in Figure \ref{fig2} represent an accumulation of such encounters.  \citet{jur09b} plots
a similar figure but including hydrogen masses in warmer DBA white dwarfs from \citet{vos07}.
Generally speaking, the hydrogen envelope masses in DBA stars are one to a few orders of 
magnitude smaller than those in the DZA, making it plausible that as helium-rich stars cool
and accrete ISM gas, the hydrogen accumulates.

If helium-rich stars gradually accumulate hydrogen from the ISM as they cool, there should
exist a significant population of cool, helium-rich white dwarfs that reveal themselves as DA 
stars, which has not yet been observed \citep{zuc03,ber01}.  Accreted hydrogen should 
perpetually float in a helium-rich star, while any metals eventually sink below detectability;
this expectation is corroborated in high-resolution spectroscopy of DB stars, where the 
metal-free DBA outnumber the DBAZ stars 15:1 \citep{vos07}.  While this ratio should
decrease in the cooler DZA temperature range as diffusion timescales and atmospheric 
opacity favor a more frequent detection of calcium, the relative number helium-rich DA and 
DZA stars should remain larger than unity for all scenarios where metals and hydrogen 
have distinct origins.  This arises because separate sources for hydrogen and metals, in 
helium atmosphere white dwarfs, imply independent probabilities, and being polluted by 
both will be the least probable.  Therefore, if one assumes that metals result from accreted 
asteroids but under-abundant hydrogen in a cool, helium atmosphere is either 1) interstellar, 
2) primordial, or 3) the result of dredge up and subsequent mixing, a larger than unity ratio 
of metal-free counterparts to the DZA stars is certain.

However, if the photospheric metals and hydrogen both have a circumstellar origin, and 
are semi-continuously delivered to the photosphere on relevant timescales, the ratio of 
helium-rich DA to DZA stars can potentially be less than unity.  There are three mixed 
helium-hydrogen atmosphere stars among the 152 cool white dwarfs thoroughly studied 
and modeled by \citet{ber01}; two are DZA stars(!), while the third has atmospheric carbon 
(where dredge up of interior carbon and helium in a previously hydrogen-rich atmosphere 
is a possibility).  While this ratio of 1:2 (or 0:2) results from small number statistics, it may 
nonetheless be telling.  The significant populations of DC and DZ stars, coupled with the 
relative lack of cool, helium-rich DA white dwarfs supports a circumstellar origin for 
hydrogen in DZA stars.

Furthermore, Occam's razor favors a single mechanism that can account for all the 
data simultaneously; hydrogen delivered together with heavy elements in water-rich 
minor planets is one such possibility.  \citet{jur09b} offer this hypothesis to explain, with 
a single pollution event, the collective facts available on the spectacularly metal-enriched, 
mixed helium-hydrogen atmosphere white dwarf GD 362.  This disk-polluted, helium-rich
star has an anomalously high hydrogen abundance at [H/He] $=-1.1$ \citep{zuc07}, so much 
that it was originally classified as DAZ \citep{gia04}.  GD 16 is a similar case with significant 
atmospheric hydrogen \citep{koe05b} and a dust disk polluting its helium-dominated 
atmosphere \citep{far09a}, while the two remaining helium-rich white dwarfs with disks 
(GD 40 and Ton 345) have little or no hydrogen \citep{jur09b}.  Therefore, the pattern of 
hydrogen abundances in DZ stars is likely a reflection of the diversity of water content in 
extrasolar planetesimals.

\section{OUTSTANDING ISSUES}

Owing to the fact that radial velocities are not reliably measurable from the calcium lines in 
the SDSS spectra of the DZ stars, some of the derived kinematical quantities are modestly 
uncertain.  Fortunately, the galactic latitudes of the 146 stars are bound within $25\degr < |b| 
< 75\degr$, and potential biases are mild at worst.  For example, although the calculation of 
$W$ velocity here is missing the third dimensional ingredient, which could potentially alter its 
sign (and hence direction), a non-zero radial velocity will tend to increase their speeds in any 
given direction [$T^2=U^2+V^2+W^2 = {v_{\rm tan}}^2 + {v_{\rm rad}}^2$].  Moreover, the focus 
of the study has been statistical in nature and the global properties of the sample are unlikely to 
change \citep{pau03,sil02}

{\em GAIA} will eventually obtain distances to, and excellent proper motions for, most if not all
the DZ and DC white dwarfs, the latter stars being immune to radial velocity measurements by 
nature (until more powerful instruments resolve any potential lines in their spectra).  Empirical
distance determinations will remove any bias introduced with the assumption of $\log\,g=8.0$, 
but again this is unlikely to be problematic from a statistical point of view \citep{kep07}.

Perhaps more important is the current understanding of how interstellar accretion occurs; this
ignorance alone can sweep up many uncertainties, just as passing stars may sweep up ISM, 
but this situation is not ideal, and empirical constraints are superior.  Although the propeller 
mechanism hypothesis may be currently broken, or minimally in need of repair \citep{fri04}, 
this does not rule out the {\em logical} possibility of preferential accretion for heavy elements 
(grains) versus gas while a star traverses a patch of ISM.  However, there are two worthwhile
considerations for any such hypothesis.  

First, based on reasonable physics in the absence of propeller-like deterrents, the infall of gas 
onto a star via the ISM should proceed at a rate much higher than the analogous accretion of 
grains (\S3.4).  Two additional factors should further increase the accreted volatile-to-refractory 
element ratio: 1) the intrinsic gas-to-dust ratio in the ISM is around 100:1, and 2) interstellar 
grain species should be separated further in the accretion process owing to varying evaporation
temperatures \citep{alc80}.  At present there does not exist a large number of cool white dwarf 
spectra exhibiting sufficient elements to confidently evaluate the volatile-to-refractory element 
ratios.  Obtaining such a database would be highly valuable.

Second, there is a class of main-sequence star which is thought to accrete ISM and become
polluted in the {\em opposite} sense to contaminated white dwarfs.  The $\lambda$ Bootis stars 
are a class of late B to early F stars that show nearly Solar abundances of light elements but 
are deficient in metals \citep{bas88}.  It has been suggested that their abundance ratios mimic 
the metal-depleted pattern of interstellar gas \citep{ven90}, and that accretion-diffusion of this 
volatile-enriched gas is consistent with the observations \citep{kam02,cha91}.  Since $\lambda$ 
Bootis stars are generally A-type dwarfs where convection has become important, they should 
share some basic characteristics with cool white dwarfs in the appropriate effective temperature 
range.  If the $\lambda$ Bootis stars are polluted by interstellar, metal-poor gas, then any ISM 
accretion theory for white dwarfs should be able to account for both classes of stars, and explain
specifically why preferential accretion would favor heavy elements in the case of white dwarfs, 
contrary to reasonable physical expectations.

\section{CONCLUSIONS}

The SDSS currently contains a large number of cool, metal-polluted, helium atmosphere 
white dwarfs over a relatively large volume of the local Galaxy.  These DZ stars appear to be, 
on average, atmospherically, spatially, and kinematically similar to a comparable number of 
DC white dwarfs from the same catalog, at least superficially.  A search for correlations between 
calcium and hydrogen abundances, spatial and kinematical properties among the DZ stars
yields negative results, excepting a tendency for increased hydrogen mass with cooling age 
in DZA stars.  Specifically, the metal abundances of 146 DZ stars shows no connection with 
speed relative to the LSR, nor with height above the Galactic disk.

The super-solar [Ca/H] ratios and lower limits are as expected for the DZ spectral class, but 
nonetheless difficult to understand as accretion of matter which is over 97\% hydrogen and 
helium.  The substantial fraction of the DZ sample which sits at $|z|>100$\,pc is difficult to
reconcile with accretion of ISM, perhaps impossible.  The total space velocities of these high 
$|z|$ stars are no slower than their counterparts within the Galactic gas and dust layer, and 
hence no more amenable to accretion of rogue ISM patches, which are unlikely to exist in
any case.  These 96 stars have spent 3 to 40\,Myr above the 100\,pc half-width scale height
of ISM, with 91 spending at least 6 Myr, the latter time period sufficient to drive their metal
abundances down by a factor 400.

The calcium masses currently residing in the convective regions of the DZ stars are near 
10$^{20}$\,g on average.  If one assumes the calcium represents 1.6\% of the total mass 
(equivalent to the bulk Earth; \citealt{all95}) of heavy elements in the stellar envelope, then a 
typical DZ star contains a mass equivalent to a 150\,km asteroid.  This is the {\em minimum} 
mass of accreted heavy elements.  If a large percentage of DZ stars have retained metals from 
a past encounter within a few to several Myr diffusion timescales, then the present envelope
masses of heavy elements represent only a fraction of the accreted masses and invoking total
captured masses as large as the Moon.

Simply taking the 146 DZ stars out of 4193 DR4 white dwarfs with $T_{\rm eff}<12\,000$\,K, 
one can say that at least 3.5\% of white dwarfs appear to be polluted by circumstellar matter, 
the remains of rocky planetary systems.  This translates directly into a similar lower limit for 
the formation of terrestrial planets at the main-sequence progenitors of white dwarfs, primarily
A- and F-type stars of intermediate mass.  While this fraction is likely to be significantly higher
\citep{zuc03}, it is difficult to quantify without a commensurate examination of all the cool SDSS 
white dwarfs, which is beyond the scope of this paper.  The appearance of hydrogen in DZA 
stars suggests a common origin for both heavy elements and hydrogen, and indicates DZA 
star pollution by water-rich minor planets may be semi-continuous on Myr timescales.  This 
picture can be tested by searching for sensitive features of oxygen absorption in hydrogen-
 and metal-polluted white dwarfs.

\section*{ACKNOWLEDGMENTS}

The authors thank an anonymous referee for helpful comments which improved the quality of 
the manuscript.  Special thanks go to N. Dickinson, S. Preval, and D. Harvey for assistance in 
mining the SDSS database.  The data presented herein are part of the Sloan Digital Sky Survey, 
which is managed by the Astrophysical Research Consortium for the Participating Institutions 
(http://www.sdss.org/).

\label{lastpage}

\end{document}